\documentclass[3p,12pt]{elsarticle}
\usepackage{graphicx}
\usepackage{epstopdf}
\usepackage[%
breaklinks%
,colorlinks%
,linkcolor=red
,anchorcolor=blue%
,pagecolor=blue%
,citecolor=blue%
,bookmarks=false%
]{hyperref}
\usepackage{amssymb}
\usepackage{amsmath}

\begin{document}

\newcommand{\rmnum}[1]{\romannumeral #1}
\newcommand{\Rmnum}[1]{\expandafter\@slowromancap\romannumeral #1@}

\begin{frontmatter}

\title{Explosion prediction of oil gas using SVM and Logistic Regression}

\author[SLY]{Xiaofei Wang}
 \address[SLY]{School of Mathematical Sciences, University of
 CAS, Beijing, China}

\author[SLY,MMZ]{Mingming Zhang}
\address[MMZ]{School of Engineering, University of CAS, Beijing, 100049, China}

\author[SLY]{Liyong Shen\corref{cor}}

 \ead{shenly@amss.ac.cn}

\author[SLY]{Suixiang Gao}

 \cortext[cor]{Corresponding author}

\begin{abstract}
The prevention of dangerous chemical accidents is a primary problem of industrial manufacturing. In the accidents of dangerous chemicals, the oil gas explosion plays an important role. The essential task of the explosion prevention is to estimate the better explosion limit of a given oil gas. In this paper, Support Vector Machines (SVM) and Logistic Regression (LR) are used to predict the explosion of oil gas. LR can get the explicit probability formula of explosion, and the explosive range of the concentrations of oil gas according to the concentration of oxygen. Meanwhile, SVM gives higher accuracy of prediction. Furthermore, considering the practical requirements, the effects of penalty parameter on the distribution of two types of errors are discussed.

\end{abstract}

\end{frontmatter}


\section{Introduction}

In recent years, there were frequent occurrences of dangerous chemical accidents in industrial processes. They bring huge damages to people's  life and property. How to predict the occurrence of dangerous chemical accidents becomes current hot topic~\cite{Rasbash1980,Hiset1997}. In 2006, James I. Chang and Cheng-Chung Lin \cite{Chang06tankacc} summarized all factory accidents in the 40 years before 2006, and classified them according to the causes of accidents. Among them, about 74\% accidents occurred at oil refining plants, oil port and oil tank. The number of accidents increase year by year. Since explosions of oil tanks may cause dominoes effects which will strengthen the damages of accidents \cite{CCPS}, it is necessary to predict and prevent the explosion of oil tank effectively.

The occurrences of fire and explosion accidents (include explosions of oil tank) need three conditions, that's, supply of oxygen, existence of combustibles and sources of kindling. If one element is controlled, combustion is impossible. According to the principles above, people naturally used to control the sources of kindling to prevent occurrences of fire/explosion accidents in the past. But since 1969, explosions of three large oil tanks happened in succession. By deep investigation and study, it comes to the conclusion that all above explosions were caused by static electricity in washing cabin \cite{Wangni2009}. From then on, people get to know a new kind of kindling-static electricity, which can hardly be removed or controlled. Then people turned to study to prevent explosion by controlling another element- oxygen.

\begin{figure}[h!]\centering
\includegraphics[scale=0.75]{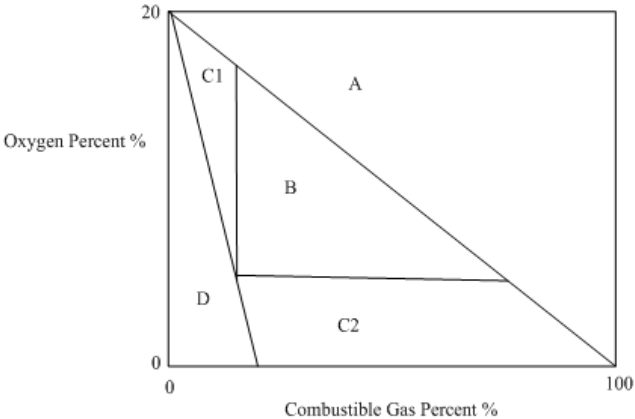}
\caption{Coward explosive triangle aera A: impossible mixture B: explosive C1: not explosive(explosive if mixed more $HC$) C2: not explosive(explosive if mixed more air) D: non-explosive}
\label{expreg}
\end{figure}

As is well known, the explosion occurs only when the concentration ratio of hydrocarbon and oxygen reaches to some certain range. Recently, there have been a lot of methods to obtain the range of explosion. In 1952, Coward and Jones \cite{CowardJones52} gave a fast and simple method to get the possibility of explosion of mixed gas. Although this method is widely used, there is a restriction that the density of reactants should be known. Jian-Wei Cheng and Sheng-qiang Yang \cite{ChengYang2011} improve the range, but the method needs the concentration of each component in the mixed gas.( see Figure~\ref{expreg}). However, in the practical oil storage and production, the compositions and densities are influenced by many factors, such as, oil production purity, temperature and pressure. And in practice, the components in the mixture are unknown, neither the densities.

With the development of artificial intelligence, nowadays the techniques of data mining are widely used into every field of chemical industry. Especially, the techniques like artificial neural network, genetic algorithm and fuzzy set theory have a lot of applications in equipment failure detection (\cite{EvsukoffGentil2005} \cite{Garcia2006}). Since our goal is to accurately predict whether explosion of oil gas happens or not, and Support Vector Machine (SVM) and Logistic Regression (LR) are efficient to tackle the classification and prediction problems(\cite{Vapnik1999, YLiuChen2007}), we decide to use the two to make prediction of oil gas explosion.

Artificial intelligence methods can improve accuracy of prediction and reduce man-made interference in which the selections of proper learning algorithms are very important. In recent years, SVM \cite{Vapnik1999} is considered as an efficient learning algorithm about pattern classification \cite{YLiuChen2007}, and the classification accuracy of SVM is better than other methods \cite{PloatGunes2008}. But the explicit expression of SVM is complex, so we can take another classification method-logistic regression \cite{WangJichuan2001} into consideration. LR has a clear explicit expression, thus it can be used to identify explosion intervals under different concentration of oxygen.

The structure of the paper is as follows. In Section~\ref{sectoilgas}, we introduce the component of oil gas, resource of data. In Section~\ref{svmlr}, we introduce some basic knowledge about SVM and logistic regression. Some experiment results are performed in Section~\ref{results}, the effect of penalty factor in SVM and comparison of two methods are also considered. Finally, we summarize the paper in Section~\ref{discuss}.

\section{Oil gas explosion}\label{sectoilgas}
In this section, we briefly give some preliminary knowledge about oilcan explosion, and introduce the explosion experiments, where the practical data are from a fire company.

\subsection{Preliminary knowledge of oilcan explosion}
The three essential requirement of oil gas explosion are:(1) Oxygen; (2) Combustibles; (3) kindling. As static electricity has become a not easy-controlled factor, we tend to control the concentration of oxygen to prevent explosion of combustion.

The limit of explosion is affected by a lot of factors, such as, temperature, oxygen content, inert medium, pressure, container, concentration of reactants. Reactants mainly contain $HC$ compound, oxygen. As the explosion occurs only when the concentration of hydrocarbon (oil gas) and oxygen reaches some mixed ratio~\cite{Wangni2009}. we establish a model to predict explosion with experimental data by data mining technique, based on the relationship between the concentration of hydrocarbon and oxygen. The prediction model is suitable for general oil gas without knowing densities.


\subsection{Source of data}
\begin{figure}[h!]\centering
\includegraphics[scale=0.6]{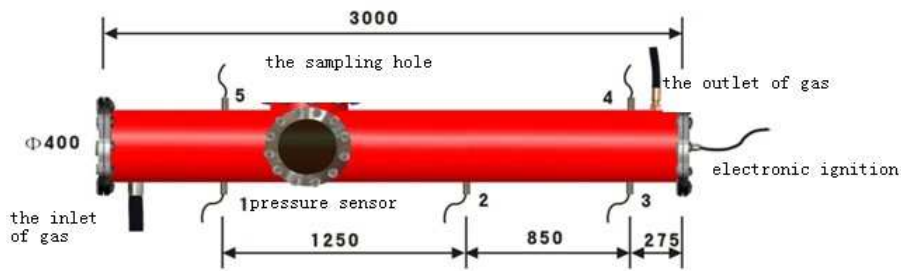}
\caption{Experiment Device}
\label{device}
\end{figure}

The data are recorded in specific explosion experiments, including the explosion pressure and explosion situation. The experiments were conducted in special closed pipe at normal atmosphere pressure and room temperature, the device is showed in Figure~\ref{device}.

Oil gas is a mixture and the concentration of each component is unknown, we consider the total concentration of $HC$, the concentration of $O_2$ and the concentration of premixed gas-mainly $CO_2$ respectively. During each experiment, the concentration of $O_2$ is controlled by injecting $CO_2$, and the densities of reactants are detected at three different points and the explosion pressure are measured at five different points by sensors, the average concentration is taken as the concentration of reactants in one experiment, so is the maximum pressure. When the data are used to predict whether explosion occurs or not, we transform them into $0$ or $1$ in LR and $-1$ or $1$ in SVM. In both of them, $1$ corresponds to the explosion, others represent no explosion. The statistic data is showed in following tables. Table~\ref{exdata} lists one experiment result in which explosion occur and Table~\ref{uexdata} lists another experiment that explosion does not occur. After taking several experiments, we get $58$ groups of data. Finally we have $58$ items and $5$ variables. Since $O_2$/$HC$ has a huge effect in reaction, so we also take it as a variable.

\begin{table}[h!]
\centering
\caption{An example of explosion data}
\label{exdata}
\newsavebox{\tablebox}
\begin{lrbox}{\tablebox}
\begin{tabular}{|c|c|c|c|c|c|c|c|c|c|c|}
  \hline
   & \multicolumn{4}{|c|}{The data before ignite} & \multicolumn{4}{|c|}{The data after ignite} & \multicolumn{2}{|c|}{Pressure(\%)} \\
  \hline
  Meas. & $HC$ & $O_2$ & $CO$ & $CO_2$ & $HC$ & $O_2$ & $CO$ & $CO_2$ & P1 & 0.080\\
  \hline
  \Rmnum{1} & 1.88 & 15.6 & 0.05 & 18.3 & 0.69 & 0.9 & 10.00 & 20.0 & P2 & 0.080 \\
  \hline
  \Rmnum{2} & 1.85 & 15.5 & 0.05 & 18.3 & 0.73 & 0.8 & 10.00 & 20.0 & P3 & 0.080 \\
  \hline
  \Rmnum{3} & 1.78 & 15.6 & 0.04 & 18.5 & 0.65 & 0.3 & 10.00 & 20.0 & P4 & 0.080 \\
  \hline
  Average & 1.84 & 15.6 & 0.05 & 18.4 & 0.69 & 0.7 & 10.00 & 20.0 & P5 & 0.080\\
  \hline
\end{tabular}
\end{lrbox}
\resizebox{0.8\textwidth}{!}{\usebox{\tablebox}}
\end{table}

\begin{table}[h!]
\centering
\caption{An example of unexplosion data}
\label{uexdata}
\newsavebox{\tableboxtwo}
\begin{lrbox}{\tableboxtwo}
\begin{tabular}{|c|c|c|c|c|c|c|c|c|c|c|}
  \hline
   & \multicolumn{4}{|c|}{The data before ignite} & \multicolumn{4}{|c|}{The data after ignite} & \multicolumn{2}{|c|}{Pressure(\%)} \\
  \hline
  Meas.& $HC$ & $O_2$ & $CO$ & $CO_2$ & $HC$ & $O_2$ & $CO$ & $CO_2$ & P1 & 0.000\\
  \hline
  \Rmnum{1} & 1.81 & 15.3 & 0.60 & 13.2 & 1.81 & 15.2 & 0.60 & 13.2 & P2 & 0.000 \\
  \hline
  \Rmnum{2} & 1.82 & 15.3 & 0.62 & 13.3 & 1.81 & 15.2 & 0.62 & 13.2 & P3 & 0.000 \\
  \hline
  \Rmnum{3} & 1.81 & 15.4 & 0.62 & 13.4 & 1.83 & 15.2 & 0.62 & 13.4 & P4 & 0.000 \\
  \hline
  Average & 1.81 & 15.3 & 0.61 & 13.3 & 1.82 & 15.2 & 0.61 & 13.3 & P5 & 0.000\\
  \hline
\end{tabular}
\end{lrbox}
\resizebox{0.8\textwidth}{!}{\usebox{\tableboxtwo}}
\end{table}

\section{ Implemental procedure of SVM and Logistic Regression}\label{svmlr}

In this section, we introduce the detailed procedure of explosion prediction using SVM and LR methods.
\subsection{Data preprocessing}

Real world data is generally incomplete, noisy, and inconsistent. After identifying input and output variables, which are introduced in Section~\ref{sectoilgas}, data need to be preprocessed. In the data preprocessing, the noisy and incomplete data are removed. And inconsistencies are corrected in the data.


For further considerations, different monitoring attributes have different scales. We need to normalize all attributes values into the same scale to avoid the influence of scales. All values of attributes are normalized to the interval [-1, 1] by using the Eq.~(\ref{Eq1}). In this equation, suppose we have $n$ samples, $v_{ij}$ is the value of $i$-th attribute of $j$-th object and $v_{ij}'$ is the normalized value. $\max{j}=\max{\{v_{1j},v_{2j},\ldots,v_{nj}\}}$, $\min{j}=\max{\{v_{1j},v_{2j},\ldots,v_{nj}\}}$. In summary, data preprocessing techniques can improve the quality of the data, so that improve the accuracy and efficiency of data mining process.

\begin{equation}
\label{Eq1}
v_{ij}'=\frac{2}{\max{j}-\min{j}}v_{ij}-\frac{\max{j}+\min{j}}{\max{j}-\min{j}}
\end{equation}

\subsection{Feature selection}
In this subsection, we use some methods to select the key features. Some knowledge of chemical industry is used to select the key features.

In the logistic regression, we use the generalized linear regression. The correlation between variables in data mining and whether the oil and gas will explode only relaying on the original variable, we expand the original data to quadratic and proportional items. However, there will be much more variables if we expand the original variables, so we need to reduce the dimension of the variables. Typically, there are two kinds of algorithms to reduce the feature space in classification. The first one is feature selection which is to select a subset of most representative feature from the original feature space. The second algorithm is feature extraction which is to transform the original feature space to a smaller one to reduce the dimension. Although feature extraction can reduce the dimension of feature space greatly compared with feature selection, the transformed smaller feature space cannot be explainable. In the test, we use the first method of feature selection; we use the all subset selection and Lasso~(\cite{lassoTi}) combined to the criterion of AIC, BIC. However, the experimental result is not good, and the difference between the original variable and variable selected by the criterion above is slight.

Due to the obvious characteristics of the variable and the fact that the number of the variable is few,  general methods of feature selection are not effective. In this paper, we use the knowledge of chemical industry and the role of the reactions to select the variable. In the reactions, $CO$ is nearly $0$ and it has no influence to the react. The role of $CO_2$ is to control the concentration of $O_2$. Thus in the paper, we do not consider the effect of $CO$ and $CO_2$. Then in the reactions, there are only two variables, $HC$ and $O_2$. Because the ratio between $HC$ and $O_2$ has serious effect on the reaction pressure in the oil and gas reactions, we take the ratio between $HC$ and $O_2$ as a variable. In the end, we get the variables, $HC$, $O_2$ and $O_2/HC$, denoted by $x_1$, $x_2$, $x_5$, respectively.

\subsection{Learning algorithms}
In this subsection, we introduce the procedure of learning algorithms for prediction, SVM and LR.
\subsubsection{Support vector machines}
As well known, SVM has been employed as the learning algorithm due to its superior classification ability. SVM is a supervised learning technique and it can be used for classification and regression. The main advantage of SVM include the use of kernel (no need to acknowledge the non-linear mapping function), the absence of local minima (quadratic problem), the sparseness of solution and the generalization capability obtained by optimizing the margin \cite{Cerqueira2008}.

Briefly speaking, SVM establishes a decision boundary between two classes by mapping the training data (through the kernel function) into a higher dimensional space, and then finding the maximal margin hyperplane within that space, which can be viewed as a classifier. Further introduction to SVM operations can be found in the following.

Given $n$ examples, $S=\{x_i,y_i\}_{i=1}^{n}$, where $x_i$ represents the condition attributes, $y_i$ is the class label, and $i$ is the number of examples. The decision hyperplane of SVM can be defined as $(\omega,b)$, where $\omega$ is a weight vector and $b$ a bias. Let $\omega_{0}$ and $b_{0}$ denote the optimal values of the weight vector and bias. Correspondingly, the optimal hyperplane can be written as
\begin{equation}
\omega_{0}^{T}\omega+b_{0}=0
\end{equation}

To find the optimum values of $\omega$ and $b$, it is required to solve the following optimization problem.
\begin{equation}
\begin{array}{ll}
\min\limits_{\omega,b,\varepsilon} & \frac{1}{2}\omega^{T}\omega+C\sum\limits_{i=1}^{n}\varepsilon_{i} \\
    \mbox{s.t.} & y_i\omega^{T}\phi(x_i)\geq 1-\varepsilon_{i},\varepsilon_{i} \geq 0
\end{array}
\end{equation}
Where $\varepsilon$ is the slack variable, $C$ is the user-specified penalty parameter of the error term$(C>0)$, and $\phi$ is the kernel function.

To sum up, SVM can change the original non-linear separation problem into a linear separation case by mapping input vector onto a higher feature space. In addition, several popular kernel functions are listed as Eqs.~(\ref{Eq4})-(\ref{Eq7}).

Linear kernel
\begin{equation}
\label{Eq4}
K(x_i,x_j)=x_ix_j
\end{equation}
Polynomial kernel of degree 
\begin{equation}
K(x_i,x_j)=(\gamma x_ix_j+r)^{g},\gamma>0
\end{equation}
Radial basis function
\begin{equation}
K(x_i,x_j)=\exp(-\gamma \parallel x_i-x_j \parallel^{2}),\gamma>0
\end{equation}
Sigmoid kernel
\begin{equation}
\label{Eq7}
K(x_i,x_j)=\tanh(\gamma x_ix_j+r),\gamma>0
\end{equation}

Here, $r$, $\gamma$ and $g$ are kernel parameters and they are user-defined. According the work of Hsu et al. \cite{HsuLin2006}, RBF kernel function is selected in this study. Readers can find more details about SVM in \cite{AslSetarehdan2008,LiuHuang2007,YLiuChen2007}.

\subsubsection{The penalty parameters of SVM}
When we use the model to predict the explosion, there are two categories of errors. The first one is the situation that the prediction result is 1 but the actual result is -1; the second one is the situation that the prediction result is -1 but the actual result is 1. SVM can set penalty upon the two types of error. Suppose $\omega_{1}$and $\omega_{2}$ are the penalty parameters of the two types of error, then the objective function of SVM that have different penalty parameters can be written as
\begin{equation}
\label{Eq8}
\frac{1}{2}\omega^{T}\omega+\omega_{1}\sum\limits_{\{j:y_i=1,\omega x_j+b=-1\}}\varepsilon_{j}+\omega_{2}\sum\limits_{\{m:y_m=-1,\omega x_m+b=1\}}\varepsilon_{m}
\end{equation}
In \ref{svmpenalty}, we will discuss the selection of penalty parameters in details.

\subsubsection{Logistic Regression}


The method of general least square is not appropriate, we can use nonlinear function. LR is a classic nonlinear classifier algorithm. The advantage of LR is that it can predict the probability of the explosion of oil gas~\cite{WangJichuan2001}.

Given $n$ examples, $S=\{x_i,y_i\}_{i=1}^{n},y_i \in \{0,1\}$, where $x_i$ represents the condition attributes, $y_i$ is the class label, and $i$ is the number of examples. LR model is to get the probability function
\begin{equation}
\label{Eq9}
p(y=1\mid X=x)=\frac{\exp(x'\beta)}{1+\exp(x'\beta)}=\frac{1}{1+\exp^{-g(x)}}
\end{equation}
Where $p$ is the probability of $y=1$ while the example chooses $x$ which is the conditional attribute, and $g(x)=x'\beta$ and $\beta$ is the parameter of LR~\cite{WangJichuan2001}.
If $p>0.5$, then $y=1$; if $p<0.5$, then $y=0$.

\subsubsection{Interval prediction using LR}
As the concentration of $CO$ is almost $0$, we omit its influence. Since the role of $CO_{2}$ is to control the concentration of $O_2$, we consider $O_2$ instead of $CO_2$. So in accordance with Eq.~(\ref{Eq9}), we get a probability function $p=p(HC,O_{2})$ for the concentration of $HC$ and $O_2$. From the function, we can get the probability of explosion, and also the explosive range of the concentrations of oil gas when the concentration of oxygen is fixed relatively.

\section{Computational results}\label{results}
In this section, we give some experimental results.
Table~\ref{dtshtind} gives a brief explanation of the data background, which includes size of data, number of variables and classification. There are totally $58$ groups of data used to analysis. Among them, there are five input variables and one output variable which is binary variables-(explosion, not explosion). In order to establish the model of prediction, we select $80\%$ of the data as training set, $20\%$ of data as test set. Table~\ref{sumattr} summarizes the variables in dataset. As stated in the above, we select $x_1$, $x_2$, and $x_5$ as investigative variables.

\begin{table}[h!]
\centering
\caption{The characteristics of dataset for thermal for storage tank}
\label{dtshtind} 
\newsavebox{\tableboxthr}
\begin{lrbox}{\tableboxthr}
\begin{tabular}{ccc}
  \hline
  \hline
  No. of records & input attributes & Target \\
  \hline
  58 & 5 Attributes & 1 target attribute with 2 classes \\
  \hline
  explosion:78\% &  &  \\
  not explosion:22\% &  &  \\
  \hline
\end{tabular}
\end{lrbox}
\resizebox{0.8\textwidth}{!}{\usebox{\tableboxthr}}
\end{table}

\begin{table}[h!]
\centering
\caption{The summary of attributes}
\label{sumattr} 
\newsavebox{\tableboxfour}
\begin{lrbox}{\tableboxfour}
\begin{tabular}{cccccc}
  \hline
  \hline
  Type & description & notation & Type & description & notation \\
  \hline
  Measure & $HC$ & $x_1$ & Measure & $O_2$ & $x_2$ \\
  \hline
  Measure & $CO$ & $x_3$ & Measure & $CO_2$ & $x_4$ \\
  \hline
  Ratio & $HC/O_2$ & $x_5$ & Status & explosion or not & $y$ \\
  \hline
\end{tabular}
\end{lrbox}
\resizebox{0.8\textwidth}{!}{\usebox{\tableboxfour}}
\end{table}


\subsection{Explosion intervals of LR}
\begin{figure}[h!]\centering
\includegraphics[width=0.7\textwidth]{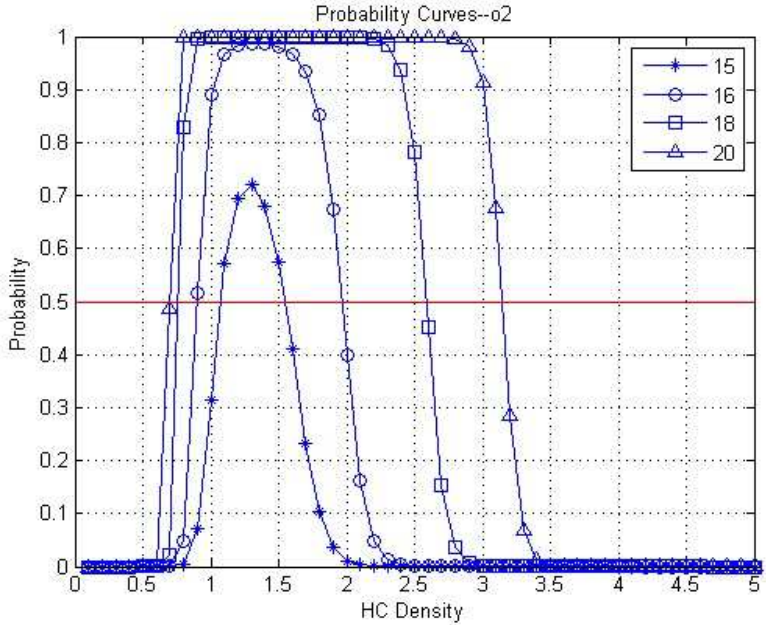}
\caption{probability}
\label{probab}
\end{figure}

\begin{figure}[h!]\centering
\includegraphics[width=0.7\textwidth]{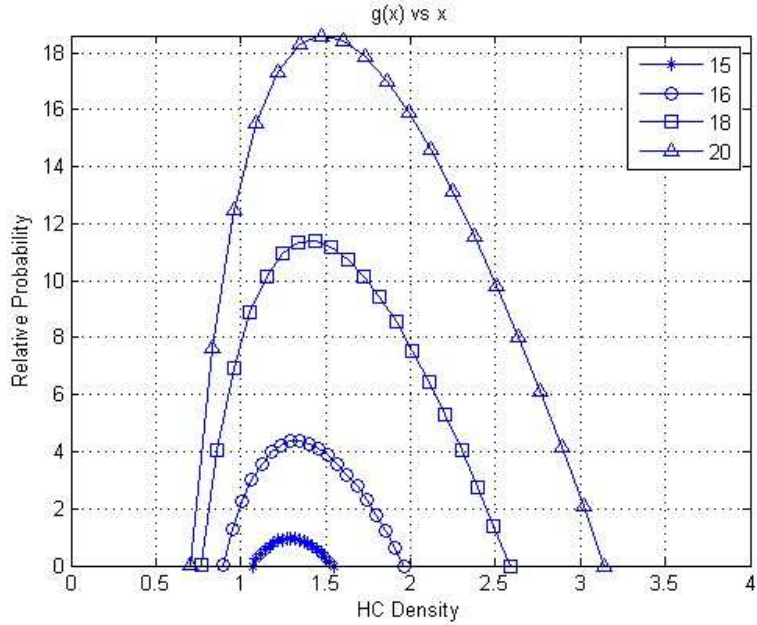}
\caption{relative probability}
\label{relpro}
\end{figure}

According to above, we use logistic regression model to obtain the explosion intervals, which are illustrated in Figure~\ref{probab} and Figure~\ref{relpro}. The $y$-axis denotes the probability of explosion with the respective $x$ under the given concentration of $O_2$, and Figure~\ref{relpro} shows the relative probability of explosion with plot of $g(x)$. The limit densities of explosion of $HC$ are summarized in Table~\ref{limex}. Here we care whether the explosion occurs and the explosion intervals. So we just compare the intervals in horizontal axis. The result performs well, it means that the logistic regression can give an intuitive explanation.

\begin{table}[h!]
\centering
\caption{The limit densities of explosion of $HC$}
\label{limex} 
\newsavebox{\tableboxintev}
\begin{lrbox}{\tableboxintev}
\begin{tabular}{ccc}
  \hline
  \hline
  Concentration of $O_2$ & the lower limit  & the upper limit \\
  \hline
  15 & 1.0668 & 1.5491   \\
  16 & 0.89729 & 1.9645  \\
  18 & 0.76653 & 2.5871 \\
  20 & 0.70066 & 3.1448 \\
  \hline
\end{tabular}
\end{lrbox}
\resizebox{0.7\textwidth}{!}{\usebox{\tableboxintev}}
\end{table}

\subsection{Comparisons between SVM and LR}
We use the popular RBF kernel in SVM, because of its good performance~\cite{CoussmentPoel2008}. RBF kernel is a uniform framework which includes other kernels and eliminate troubles caused by huge data size. As it is well known, the selection of SVM parameter has much to do with prediction accuracy. In our model, $C$ and $\lambda$ in SVM can be adjusted.

In this study, LIBSVM-3.1~\cite{ChangLin2001} is used, which is one of the most popular SVM software. In order to compare the two methods, we take one time and multi-times experiments of $v$-fold cross-validation (here, we take $v$=5). In one time experiment, by $5$-fold cross-validation, we get $5$ accuracies of prediction and the average accuracy. In multi-times experiments, we just consider the average accuracy of prediction in one $5$-fold cross-validation. The results of one time $5$-fold cross-validation are illustrated in Table~\ref{copsvmlrt}.

\begin{table}[h!]
\centering
\caption{The summary of one time 5-fold cross-validation using SVM and LR}
\label{copsvmlrt} 
\newsavebox{\tableboxfive}
\begin{lrbox}{\tableboxfive}
\begin{tabular}{ccc}
  \hline
  \hline
  Method Performance & SVM & LR \\
  \hline
  $i$-th one cross-validation & Accuracy (\%) & Accuracy (\%) \\
  \hline
  1 & 85 & 80 \\
  2 & 100 & 100 \\
  3 & 100 & 100  \\
  4 & 92 & 70 \\
  5 & 90 & 100 \\
  \hline
  Mean/6 & 93 & 90  \\
  Std & 6.54 & 13.04 \\
  \hline
\end{tabular}
\end{lrbox}
\resizebox{0.7\textwidth}{!}{\usebox{\tableboxfive}}
\end{table}

\begin{figure}[h!]\centering
\includegraphics[width=0.7\textwidth]{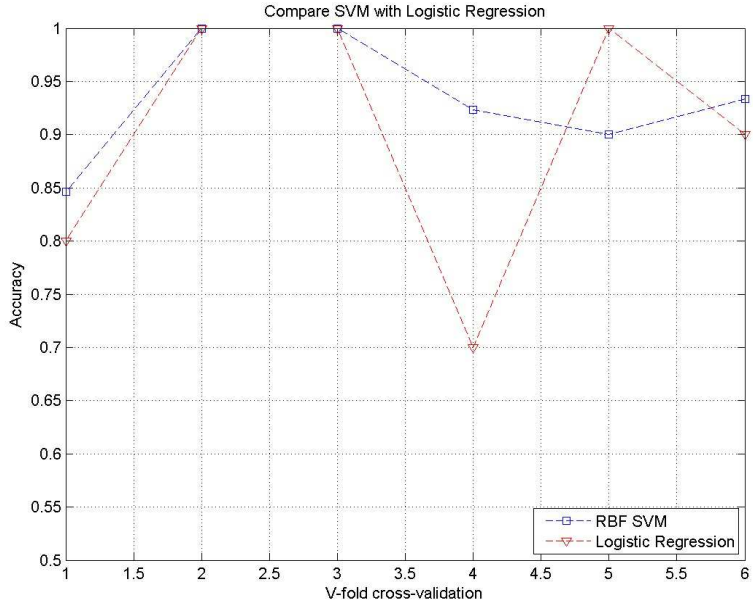}
\caption{details in one 5-fold cross validation}
\label{copsvmlrf} 
\end{figure}

The figure of one time cross-validation is showed in Figure~\ref{copsvmlrf}. (The 6-th point represents the average of the 5 points ahead.)

\begin{figure}[h!]\centering
\includegraphics[width=0.7\textwidth]{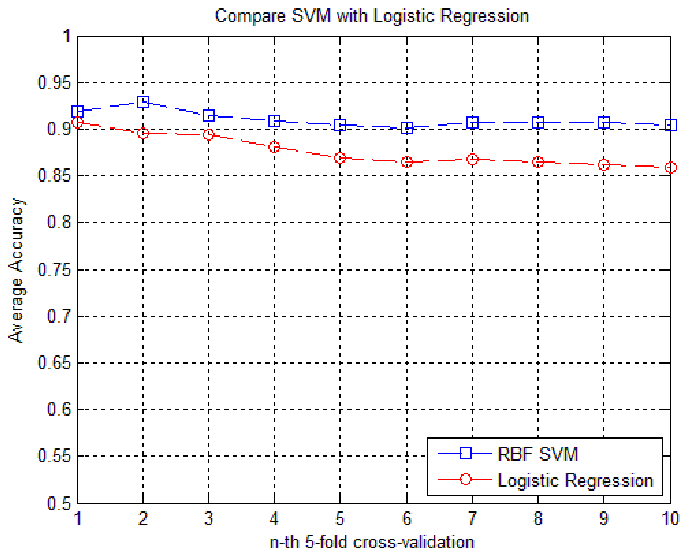}
\caption{multi-times of 5-fold cross validation}
\label{copsvmlrtf} 
\end{figure}

After one-time cross-validation, we take ten cross-validation of SVM and LR and compare the average accuracy, which are showed in Table~\ref{copsvmlrtt} and Figure~\ref{copsvmlrtf}.

\begin{table}[h!]
\centering
\caption{The summary of 10 times 5-fold cross-validation using SVM and LR}
\label{copsvmlrtt} 
\newsavebox{\tableboxsix}
\begin{lrbox}{\tableboxsix}
\begin{tabular}{ccc}
  \hline
  \hline
  Method Performance & SVM & LR \\
  \hline
  $i$-th cross-validation & Accuracy (\%) & Accuracy (\%) \\
  \hline
  1 & 92 & 91 \\
  2 & 93 & 90 \\
  3 & 91 & 89  \\
  4 & 91 & 88 \\
  5 & 90 & 87 \\
  6 & 90 & 86  \\
  7 & 91 & 87 \\
  8 & 91 & 86 \\
  9 & 91 & 86 \\
  10 & 90 & 86 \\
  \hline
  Mean & 91 & 87.6            \\
  Std & 0.94 & 1.77 \\
  \hline
\end{tabular}
\end{lrbox}
\resizebox{0.65\textwidth}{!}{\usebox{\tableboxsix}}
\end{table}

From the table~\ref{copsvmlrtt} and figure~\ref{copsvmlrtf}, it can be seen that the prediction accuracy of SVM is better than logistic regression, and logistic regression is not very stable. The minimum accuracy of SVW and LR are $91\%$ and $86\%$ respectively. And the average accuracy of the two model are $91\%$ and $87\%$ respectively. From the point view of variance, the variance of accuracy in SVM model is $0.94$, which is $1.77$ in LR model, and in one time $5$-fold cross-validation, the value are $6.54$  and $13.04$ respectively. According to both average and variance of accuracy, the prediction performance of SVM is better than that of logistic regression.

\subsection{Set penalty parameters of SVM}\label{svmpenalty}
In this section, different penalty parameters setting on the two types of error are considered.

\subsubsection{Different errors in oil gas explosion prediction}
In the procedure of prediction, it will arise two types of error. The first one is that the prediction result is 1 but the actual result is -1; the second one is that the prediction result is -1 but the actual result is 1. In our work, the first type of error means that the prediction result is not explosive but the actual result is explosive; the second type of error means on the contrary. One can find that the first error is more serious and it may lead big dangerous accident. Hence, we prefer to reduce the first error in practical predictions.
 Although these two types of error are inevitable, we can change the distribution of the types of error by adjusting parameters in SVM.

The effect of the two types of error always has a big difference. From our opinion, the first type of error is more disastrous. So we pay more attention to avoid the first type of error. According to SVM model, in Eq.~\ref{Eq8}, we set $\omega_1>\omega_2$, where $\omega_{1}$ is the penalty parameter of the first type of error and $\omega_{2}$ is that the second type of error.

From above discussion, our goal is to maximize accuracy of prediction, and meanwhile minimize the first type error, which implies that the selection of the ratio of $\omega_1$ and $\omega_2$ are quite important. In our experiments, we try to control the distribution of the two types of error by adjusting the ratio of $\omega_1$ and $\omega_2$.

\subsubsection{Balance experiments of two types of error}
\begin{figure}[h!]\centering
\includegraphics[width=0.8\textwidth]{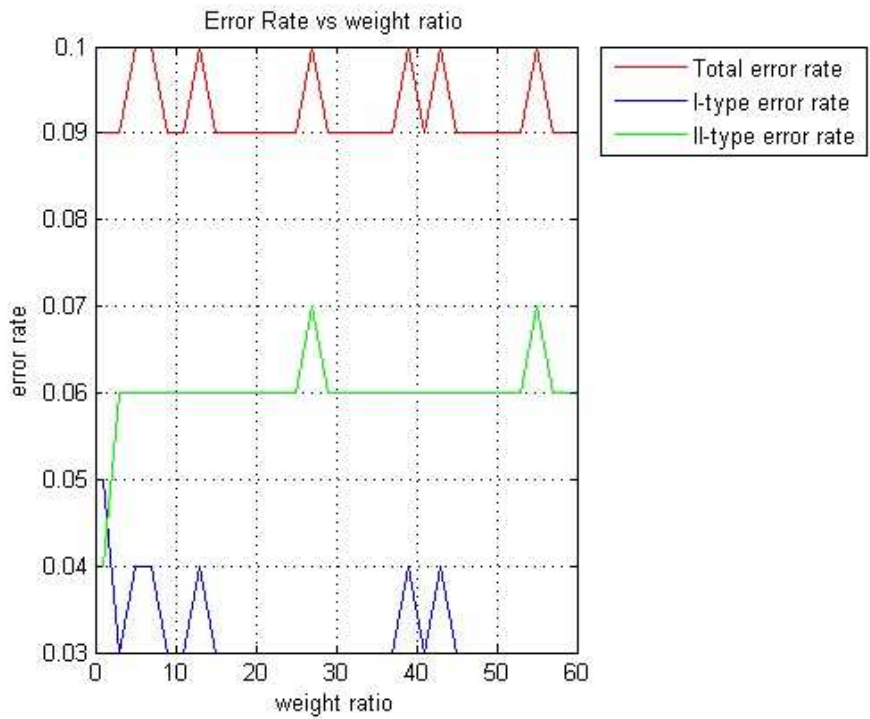}
\caption{Error Rate v.s. Weight ratio}
\label{ervswr}
\end{figure}

In this section, we choose the appropriate $\omega_1$ and $\omega_2$ to control the distribution of the two types of error. In our experiment, we construct the model when $\omega_1=\omega_2$ and $\omega_1=\gamma \omega_2$, where $\gamma\in[1,60]$ denotes the loss ratio coefficient. In this model, we take radial basis function as the kernel function. The result of the distribution of the two types of errors and the whole error rate are illustrated in Figure~\ref{ervswr} with the stepsize of $\gamma$ is 5, i.e., $\gamma=5k,k=1,\ldots,12$. As the number of data in test set is fixed, the error rate can represent the number of errors.

In Figure~\ref{ervswr}, one can find that the first error rate is larger than the second with $\omega_1=\omega_2$, which is not the case what we expect. When $\gamma>0$ increases, we can find that the first-type error rate is less than the second one. We can also find the intervals of $\gamma$ corresponding the lowest first-type error as $[5,5]$, $[8,11]$, $[15,25]$ and $[45,60]$.  The we prefer to choose the loss ratio coefficient in these intervals such that the  the whole error rate is as low as possible. In fact, we have choices of $\gamma$ such that the whole error rate is $9\%$ and the first-type error rate is only $3\%$. This result can control the distribution of the two types of errors and can achieve the goal of protecting people's life although it may loss some profit for the second errors.

\section{Discussion}\label{discuss}
In this paper, we study the explosion prediction of oil gas, in which components are unknown. We mainly concern two problems, one is to predict whether the oil gas will explode, the other is to predict the explosion interval of the given concentration of mixed $HC$.  The different statistic methods are introduced to analysis the real experimental data.  Basis the results of statistic analysis, one can predict the explosion of the given certain percentage composition of oil gas and oxygen. Precisely, we mainly introduce the methods of SVM and LR in this paper. LR method can give an explicitly probability expression. The probability expression can be used for the tanks whose data of the components of oil gas are relatively stable and having no real-time updating database. SVM method can keep learning the data with updating. Comparing to LR, SVM can get higher accuracy. And SVM applies to the area of oil tanks with the data being collected real-time. As known that it is hard to detect the percentage composition of each component in oil gas, our models concern the total percentage composition of all oil gas to predict the explosion. Thus the methods are suitable for abroad applications.

Before the explosion intervals estimating and the explosion predicting, we normalize the data and select variables for discussion. By using practical experimental data, we select the variables according to the purpose and background of the experiments. Using LR, one can obtain the explosion intervals of concentration. Considering the different levels of concentration of oxygen, we give the explosion intervals of concentration of oil gas respectively. From the computational results, we can find that the intervals of explosion are expanded to the two ends along with the increase of the concentration of oxygen. This situation just coincides with practical experience. With cross-validation in numerical experiments, the average prediction accuracy of LR can reach $87\%$, while the SVM gets more than $90\%$. These results show the applicability of SVM and LR for predicting the explosion of oil gas.

In practical security protection with some special purposes, higher accuracy of prediction for some situations is needed. Generally, there are two kinds of error in prediction. One is that the no explosion is predicted but the data is in the range of explosion actually. Another is that the explosion is predicted for the new data but it is not true. We cannot tell which is more important of the different errors in theoretical analysis. However, in practical application, the second type of error may induce oil tank fire, which leads to disastrous consequences. According to practical needs, we can restrict its occurrence as less as possible to meet the practical security requirements.

\section*{Acknowledgement}
The authors would like to thank the Professor Sanguo Zhang for his useful advice and comments on this paper. We also thank Weitelog Fire Company for the data experiments. This work is supported by National Key Basic Research Program of China under Grant No.2011CB706901.

\bibliographystyle{model1-num-names}
\bibliography{DangerChem}

\end{document}